\documentclass[sigconf]{acmart}
\usepackage{booktabs} 
\usepackage{soul}
\usepackage{bm}
\usepackage{algorithm}
\usepackage{algpseudocode}
\usepackage{comment}
\usepackage{verbatim}
\usepackage{hyperref}
\usepackage{fancybox}
\usepackage{balance}
\usepackage{graphicx}
\usepackage{epstopdf}
\usepackage{color, colortbl}
\definecolor{Gray}{gray}{0.9}

\usepackage{ifthen}
\usepackage{amssymb}
\newboolean{showcomments}
\setboolean{showcomments}{true} 
\ifthenelse{\boolean{showcomments}}
  {\newcommand{\nb}[2]{
    \fcolorbox{gray}{yellow}{\bfseries\sffamily\scriptsize#1}
    {\sf\small$\blacktriangleright$\textit{#2}$\blacktriangleleft$}
   }
   
  }
  {\newcommand{\nb}[2]{}
   
  }

\newcommand{\todocite}[1]{\smash{\fcolorbox{gray}{red}{[?]}}}

\setcopyright{rightsretained}

\acmConference[BotSE'20]{International Workshop on Bots in Software Engineering}{May 2020}{South Korea} 

\begin{document}
\title{An Exploratory Study of Bot Commits}

\author{Tapajit Dey}
\orcid{0000-0002-1379-8539}
\affiliation{%
  \institution{The University of Tennessee}
  \streetaddress{1520 Middle Dr.}
  \city{Knoxville}
  \state{TN}
  \country{USA}
  \postcode{37996}
}
\email{tdey2@vols.utk.edu}


\author{Bogdan Vasilescu}
\affiliation{%
  \institution{Carnegie Mellon University}
  \streetaddress{Forbes Avenue}
  \city{Pittsburgh}
  \state{PA}
  \country{USA}
  \postcode{15213}
}
\email{vasilescu@cmu.edu}

\author{Audris Mockus}
\affiliation{%
  \institution{The University of Tennessee}
  \streetaddress{1520 Middle Dr.}
  \city{Knoxville}
  \state{TN}
  \country{USA}
  \postcode{37996}
}
\email{audris@mockus.org}

\renewcommand{\shortauthors}{T. Dey et al.}

\begin{abstract}
Background: Bots help automate many of the tasks performed by software developers and are widely used to commit code in various social coding platforms. At present, it is not clear what types of activities these bots perform and understanding it may help design better bots, and find application areas which might benefit from bot adoption.  
Aim: We aim to categorize the Bot Commits by the type of change (files added, deleted, or modified), find the more commonly changed file types, and identify the groups of file types that tend to get updated together.
Method: 12,326,137 commits made by 461 popular bots (that made at least 1000 commits) were examined to identify the frequency and the type of files added/ deleted/ modified by the commits, and association rule mining was used to identify the types of files modified together.
Result: Majority of the bot commits modify an existing file, a few of them add new files, while deletion of a file is very rare. Commits involving more than  one type of operation are even rarer. Files containing data, configuration, and documentation are  most frequently updated, while HTML is the most common type in terms of the number of files added, deleted, and modified. Files of the type ``Markdown'',``Ignore List'', ``YAML'', ``JSON'' were the types that are updated together with other types of files most frequently. 
Conclusion: We observe that majority of bot commits involve single file modifications, and bots primarily work with data, configuration, and documentation files. A better understanding if this is a limitation of the bots and, if overcome, would lead to different kinds of bots remains an open question.
\end{abstract}

%
%


\maketitle

\section{Introduction}\label{s:intro}
Collaborative software development and social coding platforms have 
seen a surge in bot adoption in recent years~\cite{wessel2018power}. 
A quick look at the users who create the most number of commits, 
issues, and/or pull requests reveals that most of them are, in fact, 
bots~\cite{dey2020biman}.
A number of past studies have attempted to categorize the bots that 
are active in social coding platforms by their origin, purpose, and 
function (e.g.~\cite{erlenhov2019current,lebeuf2018taxonomy}), however, 
the activities of the bots have not been directly studied in existing 
literature. 

A study of the commits made by bots can give us valuable insights into 
their function and application domain, which can be useful for developers 
planning to adopt bots in their projects in terms of identifying the 
types of tasks typically handled by bots and the languages they are 
active in. 
The developers who design bots might also benefit from this knowledge 
by identifying the gaps in the current bot landscape in software 
engineering. 

Therefore, in this paper we report on an exploratory study of the 
commits created by bots in a large dataset~\cite{tapajit_dey_2020_3694401}
of Git open-source projects, introduced by~\citet{dey2020biman}.
Specifically, we investigated 12,326,137 commits from 461 popular bots 
to examine the frequency and the type of files changed by the commits, 
categorized the commits in terms of the files added, deleted, or 
modified therein, and identified which types of files frequently get 
changed together with other types of files.

Our results show that an overwhelming majority (86\%) of the bot 
commits involve only file modification, with 68\% of the commits 
changing only a single file, and only 5\% of the commits changed more 
than 10 files. 
Data, configuration, and documentation files are the most frequently 
updated ones, but HTML takes the crown in terms of the number of files 
added, deleted, and modified. 
A lot of Java files were found to be added by bots, but not modified 
frequently, so it most likely was for the purpose of archiving. 
We also noticed that files of types ``Markdown'', ``YAML'', ``Ignore List'', 
``JSON'', and ``Text'' are the ones most frequently updated with other 
types of files in the bot commits that do change more than one file. 

We have also compiled a dataset containing information about 150,633,947 
file updates by the 12,326,137 bot commits, specifically information 
about the ``blobs''\footnote{\url{https://git-scm.com/book/en/v2/Git-Internals-Git-Objects}} associated with the files before the commit, and the updated 
blobs. 
The dataset is available at: \url{https://zenodo.org/record/3699665}.

In summary, we find that:
\begin{itemize}
    \item Bots are mostly taking care of file updates, and update a small 
    number of files per commit, so designing bots with limited scope seems 
    to be the current trend in Software Engineering. 
    \item The majority of bot commits comprises frequent updates to 
    configuration, documentation, and data. 
    Developers planning to adopt bots for their projects might want to 
    consider using bots for similar tasks. 
    \item Bots seem to be more active in Web-interface-related projects, 
    since the majority of bot commits involves changes to HTML, JavaScript, 
    and JSON files. 
    Future researchers might want to investigate the reason for the popularity 
    of bots in this area, which might lead to better design of future bots.
\end{itemize}

\section{Related Work}\label{s:relwork}
Bots are regularly used in a number of areas like education~\cite{kerry2008conversational, benotti2014engaging}, e-commerce~\cite{thomas2016business}, customer service~\cite{jain2018convey}, and social media platforms~\cite{abokhodair2015dissecting}.
In Software Engineering, bots are typically used to automate a number of, often tedious and repetitive, tasks performed by software developers and teams. Wessel et
al.~\cite{wessel2018power} found 26\% of the 351 GitHub projects they studied  use bots.

In terms of bot characterization, Lebeuf~\cite{lebeuf2018taxonomy} proposed characterizing the bots by analyzing 22 facets grouped into 3 dimensions: Environmental, Intrinsic, and Interaction. Erlenhov et. al.~\cite{erlenhov2019current} focused on the 11 well-known bots that support software development, and proposed a taxonomy comprising 4 facets: Purpose (general vs. specialized), Initiation (triggered by users or system or both), Communication (how the bot communicates with other users), and Intelligence (adaptive or static). 

A novel method for detecting which commit authors are bots was proposed in Dey et al.~\cite{dey2020biman}, which looked at the author name, the commit message, and the 
files and projects associated with a commit. They compiled a dataset~\cite{tapajit_dey_2020_3694401} containing information
about 13,762,430 commits made by 461 popular bots, each of whom made more than 1000 commits.
In this paper, we used that dataset to study the individual commits in further detail.
While other studies (e.g.~\cite{dey2019patterns}) investigated the activity of human developers, to our knowledge, no study has investigated the bot commits in details.

\section{Methodology}\label{s:method}
In this section, we describe the data source and the analysis techniques used in this paper.

\vspace{-10pt}
\subsection{Data Source}
As mentioned earlier, we used the dataset~\cite{tapajit_dey_2020_3694401} to obtain a list of the bot commits. The dataset contained information about 13,762,430 commits, however, information about the exact file modification (i.e. the blobs associated with files that were updated by the commit) was not available for a few of them, so, after filtering them out, we were left with 12,326,137 commits.

Detailed information about the files updated by the bot commits were extracted using the World of Code~\cite{woc19} dataset, which is a prototype of an updatable and expandable infrastructure to support research and tools that rely on version control data from the entirety of open source projects that use Git. 
We used version Q of the dataset for the analysis described in this paper. This data (version Q) was collected on Nov 9 based on updates/new repositories identified on Oct 15, 2019. For further details, please check the WoC website.~\footnote{\url{http://worldofcode.org}}

\vspace{-10pt}
\subsection{Analysis Method}
The data in WoC is stored in the form of mappings between various git objects, and, using the APIs provided by WoC, we constructed a mapping (which we identify as \textit{c2fbb} maps) between a commit, the file(s) modified by that commit, and the old blob associated with each file before the commit, and the new blob associated with the updated version of that file. If either the old blob or the new blob associated with a file was found missing, it meant the file was added or deleted, respectively, by that commit, and if both were found present, it meant that file was modified by that commit.

We started our analysis by investigating the file types that were added, deleted, or modified by each commit. We extracted the file extensions from those files, and used the \textit{linguist}~\footnote{https://github.com/github/linguist} tool to obtain an estimated language classification based on a common open-source model. 

To find out what types of files are updated together, we used association rule mining on the file types updated in each commit using the ``arules''~\cite{arules} package in R. Since most of the bot commits were found to update only one type of file, we applied this technique only on the commits that update two or more types of files, so that the result doesn't get overwhelmed by singletons. We used a minimum support of 0.1\% and a confidence of 80\%, along with a maximum length of 10 types of files for a single association rule for the \texttt{apriori} function call.
We used the ``arulesViz''~\cite{av} package in R for visualizing the results of this analysis.

\vspace{-10pt}
\section{Result}\label{s:result}

\subsection{Shared Dataset:}
We compiled a dataset with the information about the blob updates related to each file  updated by one of the bot commits, so that other researchers may also study the bot commits. Details about the structure of the dataset, and how to use the GitHub API to extract the contents of each blob is mentioned in the description of the dataset available through the link in Section~\ref{s:intro}.

\vspace{-10pt}
\subsection{Categorizing Bot Commits}

\begin{figure}[!t]
\centering
\includegraphics[width=0.8\linewidth]{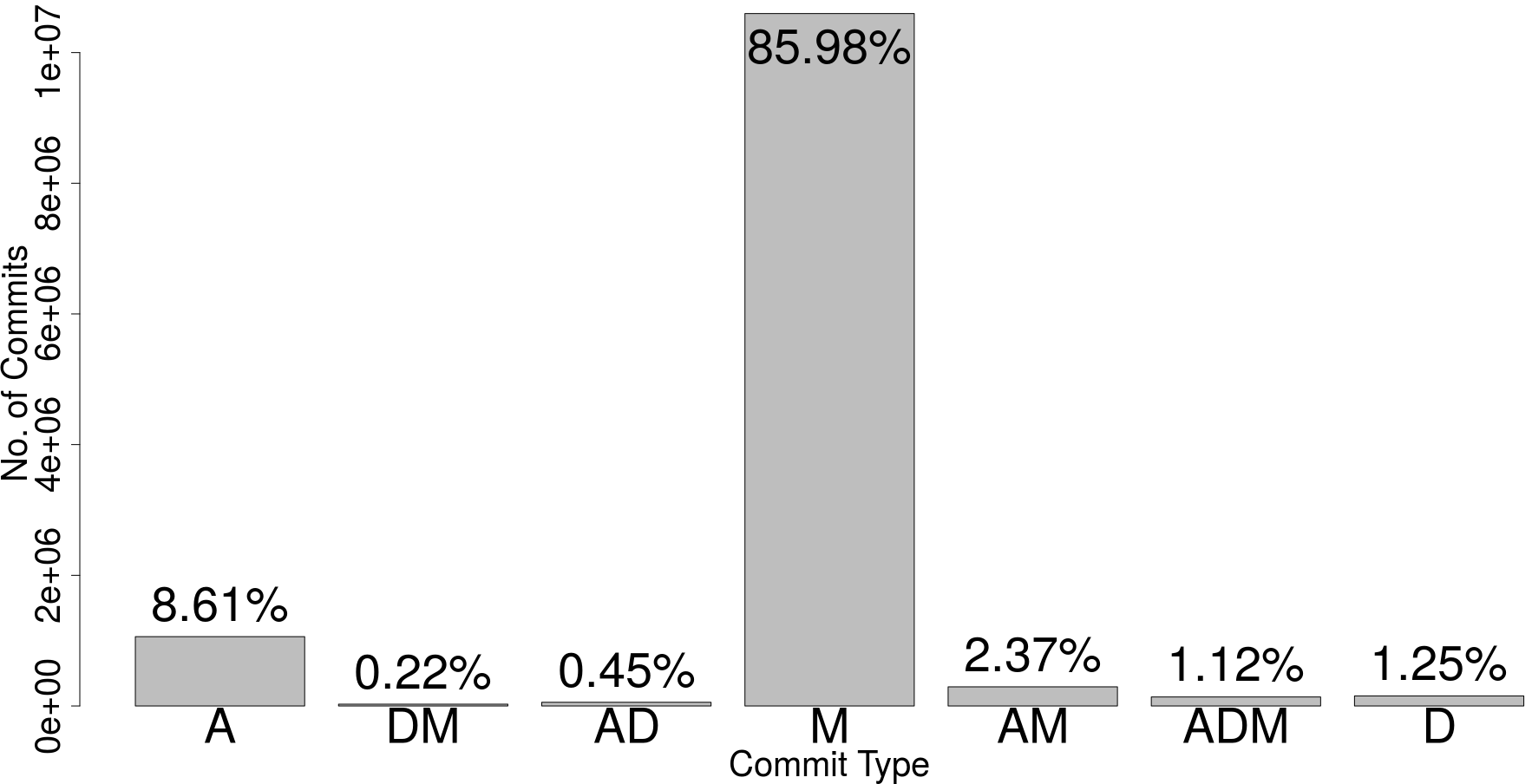}%
\caption{Distribution of the number of commits of each type, with percentage of each commit type shown on the X-axis.}
\label{fig:comtype}
\vspace{-15pt}
\end{figure}

By observing the \textit{c2fbb} maps, we identified all the files that were updated by the bot commits under consideration, along with the old and new blobs associated with the file, and could also identify which files were added, deleted, or modified by each commit (see Section~\ref{s:method}). We categorized the bot commits into 7 categories by the type of file change (addition, deletion, and modification): 
\vspace{-2pt}
\begin{enumerate}
    \item Type \textbf{A}: Commits involving only file addition.
    \item Type \textbf{D}: Commits involving only file deletion. 
    \item Type \textbf{M}: Commits involving only file modification.
    \item Type \textbf{AM}: Commits involving file addition and modification.
    \item Type \textbf{DM}: Commits involving file deletion and modification.
    \item Type \textbf{AD}: Commits involving file addition and deletion.
    \item Type \textbf{ADM}: Commits involving file addition, deletion, and modification.
\end{enumerate}
\vspace{-5pt}
The relative prevalence of each type of commit is shown in Figure~\ref{fig:comtype}, which shows majority (85.98\%) of the commits involve only file modifications (Type \textbf{M}), while more complex commits (ones with more than one type of file change: Types \textbf{AM, DM, AD, ADM}) are relatively uncommon. Interestingly, commits involving only file deletions (Type \textbf{D}) are rarer than commits where some files were added and some were modified (Type \textbf{AM}), and complex commits involving file deletions (Types \textbf{DM, AD, ADM}) are even rarer. Bots that modify files include bots like \textit{GreenKeeper} that update dependency versions for NPM packages, archival bots like \textit{AUR Archive Bot} are responsible for adding a lot of files, while a number file deletions was performed by bots responsible for checking in human commits after running tests, e.g.  instances of \textit{Travis CI bots}.

\begin{figure}[!t]
\centering
\includegraphics[width=0.8\linewidth]{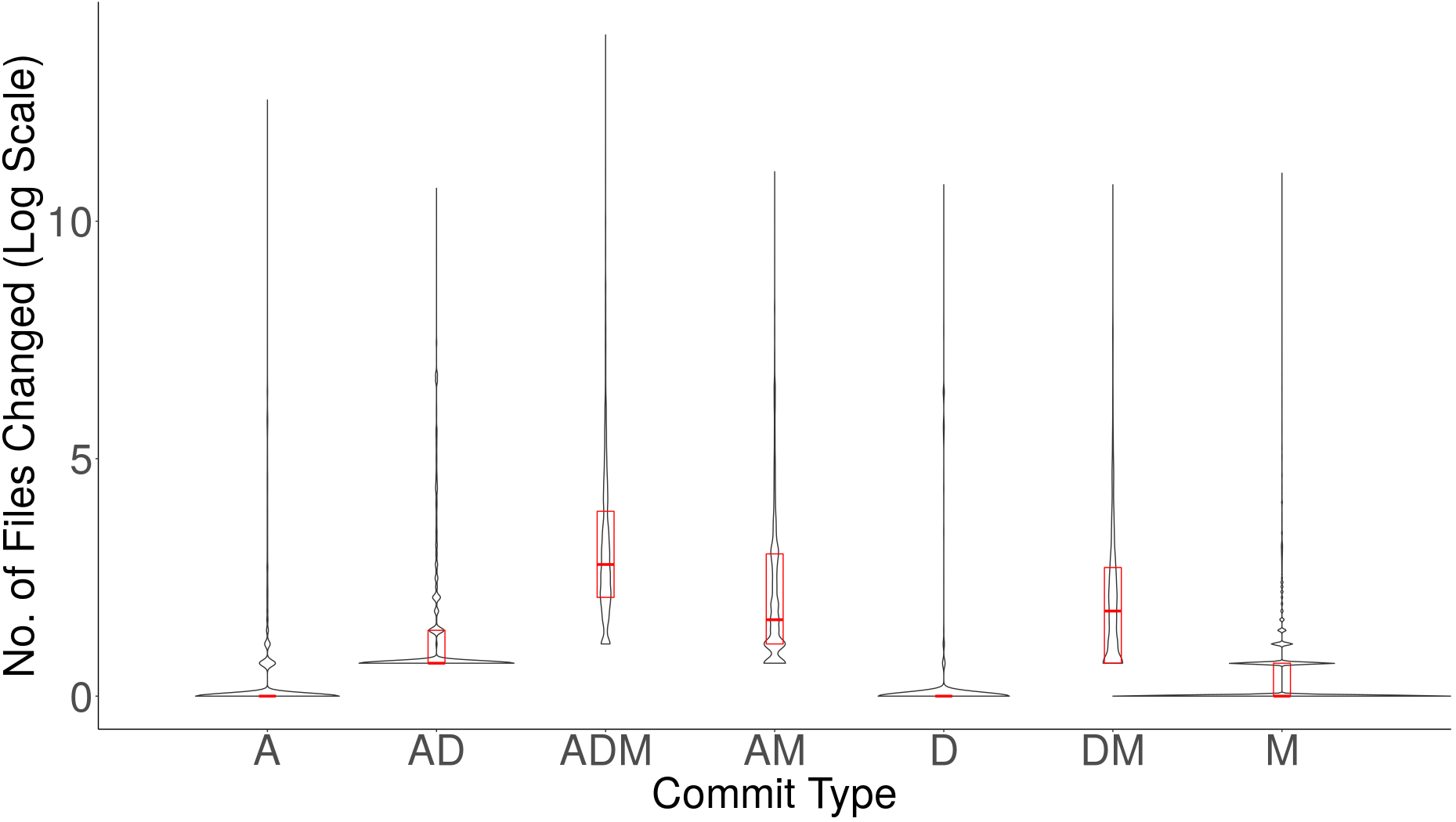}%
\caption{Distribution of the number of files modified by each type of bot commit.}
\label{fig:fltype}
\vspace{-15pt}
\end{figure}

We also looked at the commit sizes for the bot commits, i.e.\ how many files were changed by each commit. Looking at the overall picture, we observed most commits to be of very small size, with the median no. of files changed being 1, but the maximum number of files changed by a commit was observed to be a whooping 1,113,522. We also investigated the distribution of the number of files changed by commits of different types, which is shown in Figure~\ref{fig:fltype} using a violin plot, with the median and 1st and 3rd quantile values shown in a (red) crossbar plot. We observed a highly skewed distribution of the number of files changed for all commit types, however, the more complex commit types were observed to have relatively larger median number of files changed.

While Figure~\ref{fig:fltype} shows that most commits change very few files, we were curious about how many different file types are changed by the commits to get a sense of the degree of heterogeneity in the commits. We found that 10,270,857 (83\%) commits have changed only one type of file, while the maximum number of different file types changed by a bot commit was found to be 121. 

We can infer from these results that bots are primarily designed for simpler updates involving modification to a few files, which leads us to believe that they have a limited scope, so developers interested in adopting bots might want to consider a similar focus for their application.

\begin{figure}[!t]
\centering
\includegraphics[width=0.75\linewidth]{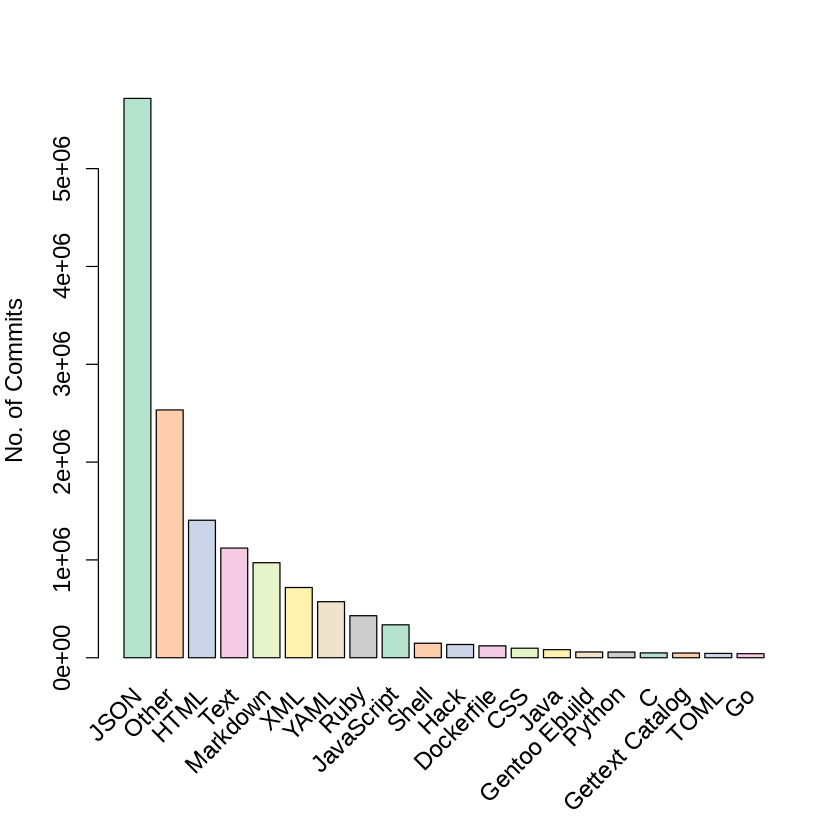}%
\vspace{-15pt}
\caption{Different file types updated by bot commits.}
\label{fig:flall}
\vspace{-15pt}
\end{figure}

\begin{figure*}[t!]
        \centering
        \begin{minipage}[c]{0.26\textwidth}
            \centering
            \noindent\includegraphics[width=\textwidth]{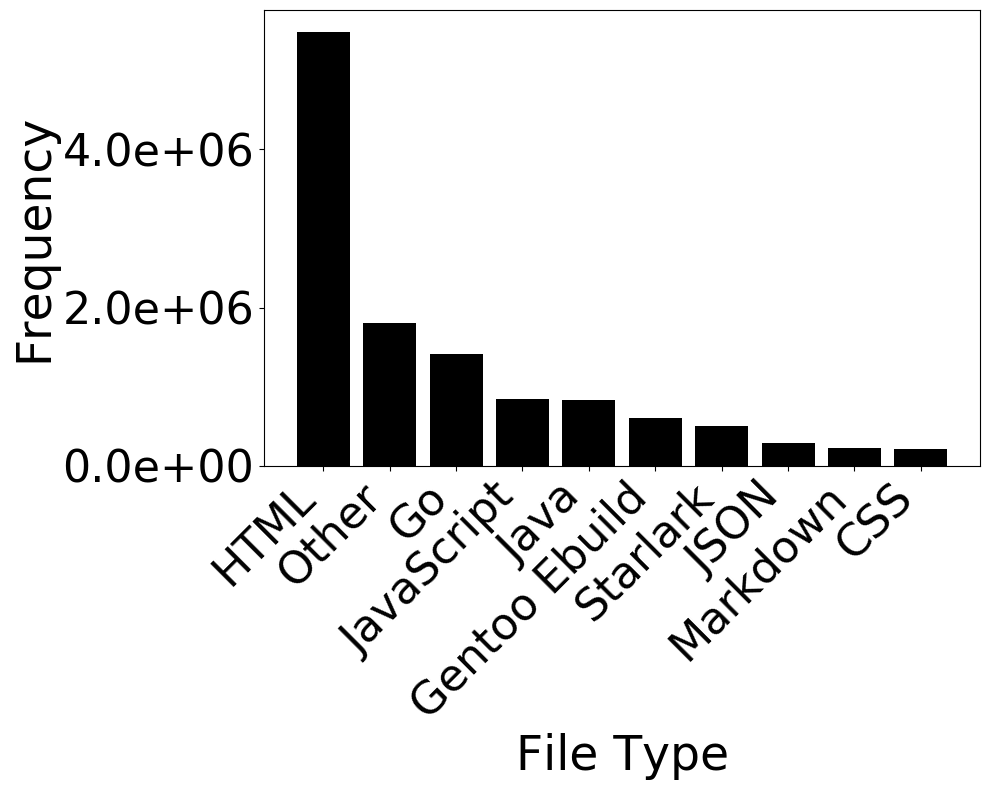}
            a
        \end{minipage}
        \begin{minipage}[c]{0.26\textwidth}
            \centering
            \noindent\includegraphics[width=\textwidth]{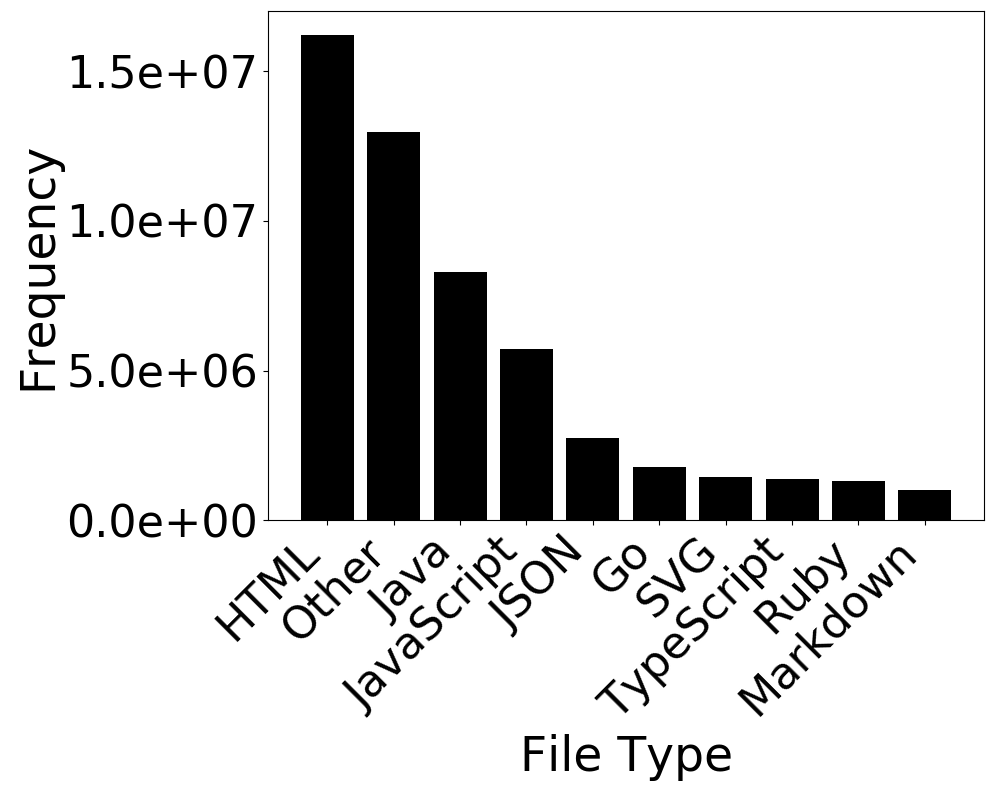}
            b
        \end{minipage}
        \begin{minipage}[c]{0.26\textwidth}
            \centering
            \noindent\includegraphics[width=\textwidth]{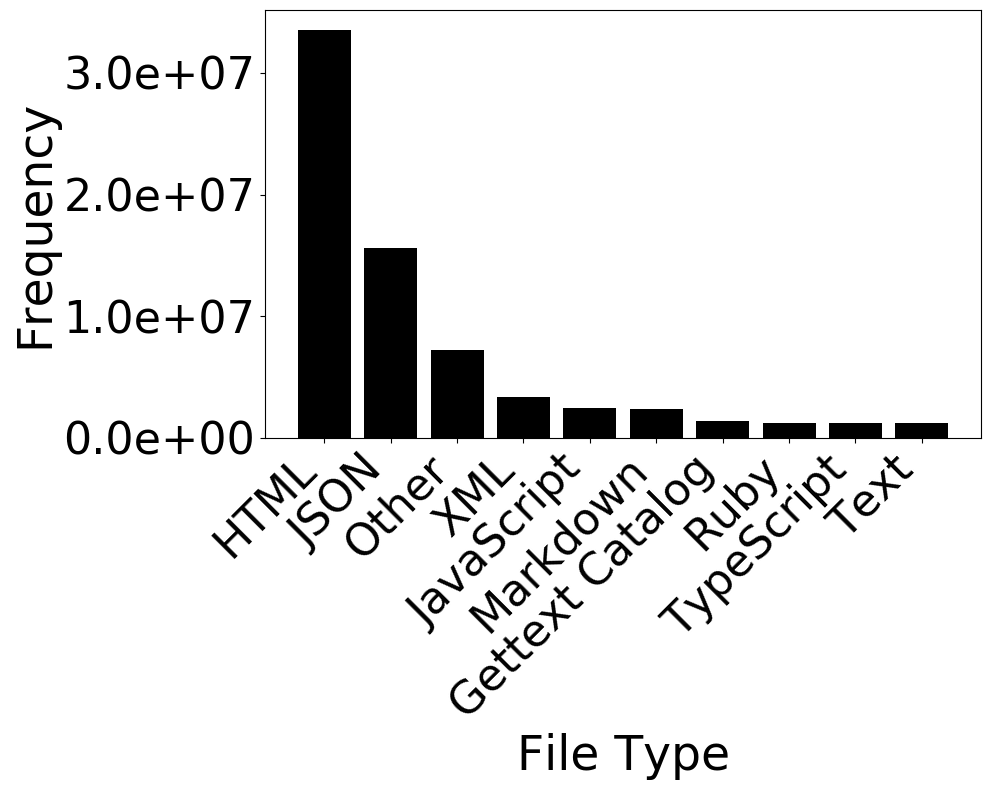}
            c
        \end{minipage}
        \vspace{-5pt}
        \caption{File Types changed by bots: (a): Top 10 deleted, (b)Top 10 added, (c):Top 10 modified.}
        \label{fig:addelmod}
\vspace{-10pt}
\end{figure*}







\vspace{-10pt}
\subsection{Types of Files changed by Bot Commits}
We observed that out of the 150,633,947 instances when a file was changed by one of the bot commits,  50.1\%,  40.9\%, and 9\% are of file modification, file addition, and file deletion respectively. 

After inferring the language of the files using the GitHub Linguist tool,~\footnote{https://github.com/github/linguist}  we decided to observe what types of files are added, deleted, and modified most frequently by bot commits. Any file type that was not defined in the Linguist tool (e.g. files with extensions like \texttt{.icloud, .AURINFO} etc.) were categorized as ``Other'', which was primarily comprised of files specific to a particular software. Figure~\ref{fig:flall} shows the top 20 file types that are updated by different bot commits, with the number of commits that updated a particular file type in the Y-axis. We see that JSON files are updated by the most number of commits, followed by ``Other'' and HTML files. Overall, we see configuration, documentation, and other data related files are updated more frequently than code related files. Ruby, JavaScript, and Shell are the top three languages updated most frequently. Different types of commits showed very similar distributions to the overall situation for the file types in terms of the number of commits that updated it, so we do not show those plots separately.

We also looked at the number of files of different types that are updated by the bot commits, and we observed that HTML was the most frequently deleted, added, and modified file type, as shown in Figure~\ref{fig:addelmod}. Java files are among the most frequently added file type, Go files are among the most frequently deleted ones, and JSON files are the second most frequently modified file type. Comparing what we see from Figure~\ref{fig:flall}, we can infer that the average number of HTML files updated per commit is much higher than the average number of JSON files updated per commit, since Figure~\ref{fig:flall} shows the number of commits that updated a particular file type, whereas Figures~\ref{fig:addelmod} show the total number of files (of each type) updated. It also seems from our results that most of the bot updates involve updates to Web interface related files, so bot designers might want to look into expanding automation to other domains as well.

\begin{figure}[!t]
\centering
\includegraphics[width=0.82\linewidth]{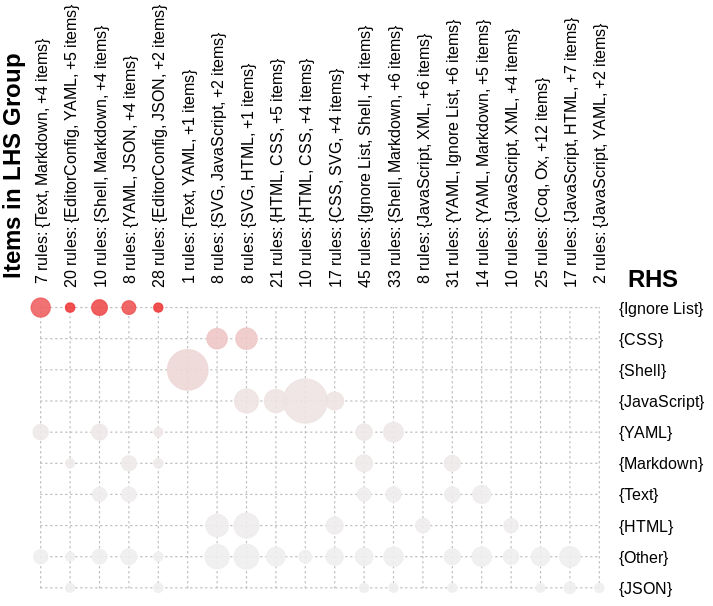}%
\caption{Visualizing grouped association rules} 
\label{fig:arAll}
\vspace{-20pt}
\end{figure}

\vspace{-10pt}
\subsection{File Types frequently updated with others}


As mentioned earlier, only 17\% of all the bot commits change more than one type of file, so investigating what types of files are updated together with other types can give us some insight about these relatively infrequent heterogeneous commits. We applied association rule mining on these commits to address this question, as mentioned in Section~\ref{s:method}, and obtained a set of 323 non-redundant association rules, with the minimum, maximum, and median sizes of the rule lengths being 2,6, and 4 respectively. The support for the rules varied between 0.001 and 0.006, with a confidence range between 0.802 and 0.998. The lift for the  rules was observed to be between 1.78 and 378.93, with a median of 9.79.

We show a grouped representation of all the rules in Figure~\ref{fig:arAll}, where similar rules are grouped together for ease of representation. The size of each circle in the figure corresponds to the support associated with that rule, and the intensity of the color red corresponds to the lift associated with the rule. An interesting observation from this figure is that rules with files of type ``Ignore List'' (e.g. \texttt{.gitignore}) have very high values of lift (these are also the rules with most confidence), i.e. updates to ``Ignore List'' file types occur more frequently in the commits that update other file types (as shown in Figure~\ref{fig:arAll}) as well. The file types on \textbf{RHS} column in the figure are the ones that appear more frequently with other file types than by themselves (since the lift for all the rules we have is positive). We also see that files of type ``Other'' were associated with most of the rules, and rules with ``Shell'' and ``JavaScript'' file types in \textbf{RHS} have the highest support. We observe that most file types in the rules are related to data, configuration, documentation, and web-design, similar to what we observe overall, and file types that are updated together tend to be complimentary (e.g. HTML with JavaScript and CSS).

\begin{table}[ht]
\centering
\caption{Top 5 Association Rules by lift}
\label{t:arlift}
\vspace{-10pt}
\resizebox{\linewidth}{!}{%
\begin{tabular}{p{3cm}clccc}
  \hline
\textbf{LHS} &  & \textbf{RHS} & \textbf{support} & \textbf{confidence} & \textbf{lift} \\ 
  \hline
\{EditorConfig, Markdown, Other, YAML\} & =$>$ & \{Ignore List\} & 0.001 & 0.99 & 378.93 \\ 
\{EditorConfig, JSON, Other, YAML\} & =$>$ & \{Ignore List\} & 0.001 & 0.99 & 378.74 \\ 
\{EditorConfig, Other, YAML\} & =$>$ & \{Ignore List\} & 0.001 & 0.99 & 378.26 \\ 
\{EditorConfig, JSON, Markdown, YAML\} & =$>$ & \{Ignore List\} & 0.001 & 0.99 & 378.26 \\ 
 \{EditorConfig, JSON, Markdown, Other\} & =$>$ & \{Ignore List\} & 0.001 & 0.99 & 378.09 \\ 
   \hline
\end{tabular}
}
\vspace{-20pt}
\end{table}

We show the top 5 association rules with highest lift values in
Table~\ref{t:arlift}. For all 5 rules, ``Ignore List'' file type was in the 
\textbf{RHS}, and ``EditorConfig'' was one of the types in \textbf{LHS}. The confidence for these rules was found to be very high ($0.99$), and the associated support values were $0.001$ for all of them.


\vspace{-10pt}
\section{Limitations}\label{s:limit}
We inferred the types of files directly from their names, instead of looking into the contents of the files, which comes with an obvious risk of error. In addition, we faced some challenges while using the Linguist tool for file classification, since some file extensions were found to be linked with multiple file types (e.g. ``.gs'' files are associated with GLSL, Genie, Gosu, and JavaScript file types). We addressed this problem by adding an entry to all possible types when we encountered such cases, which introduces some error in our result. However, this scenario is not very common (less than 1\%), so it should not cause too much of a problem.

We also only focused on the commits by 461 bots that made over 1000 commits each, out of possibly thousands of bots that make code commits, so our results may not generalize for the overall bot population. 

\vspace{-10pt}
\section{Future Work}\label{s:future}
In future, we would like to extend our study of bot commits by looking into the text diffs between the old and the updated blobs associated with each file changed by each commit, which would give us further insight into what exact changes are made by the bots in their commits. Furthermore, we would like to address the issue of multiple IDs related to a bot, by using the methodology proposed in~\citet{fry2020dataset}, and also investigate how the presence of bots affect the popularity~\citet{dey2018software} of a software and, in turn, affect the number of issues~\citet{dey2018modeling,dey2020deriving} and pull request acceptance~\citet{dey2020pull}.

\vspace{-10pt}
\section{Conclusion}\label{s:conclusion}    
In this paper, we investigate the commits created by bots, categorize the commits based on the type of file operation they perform and what types of files they change to understand the types of works performed by bots at present in the context of software engineering, and insights from this study might be valuable in understanding the strengths and limitations of the bots presently active, and can, in future, lead to better design and wider adoption of bots.

\balance
\bibliographystyle{ACM-Reference-Format}
\bibliography{sigproc} 

\end{document}